\documentclass[aps,prl,reprint,superscriptaddress]{revtex4-1}
\usepackage{graphicx,xcolor,soul}
\usepackage{amsmath}
\usepackage{float}
\begin{document}


\title{Anomalous sequence of quantum Hall liquids revealing a tunable Lifshitz transition in bilayer graphene}


\author{Anastasia Varlet}
\email[]{varleta@phys.ethz.ch}
\homepage[]{www.nanophys.ethz.ch}
\affiliation{Solid State Physics Laboratory, ETH Z\"{u}rich, 8093 Z\"{u}rich, Switzerland}

\author{Dominik Bischoff}
\affiliation{Solid State Physics Laboratory, ETH Z\"{u}rich, 8093 Z\"{u}rich, Switzerland}

\author{Pauline Simonet}
\affiliation{Solid State Physics Laboratory, ETH Z\"{u}rich, 8093 Z\"{u}rich, Switzerland}

\author{Kenji Watanabe}
\affiliation{Advanced Materials Laboratory, National Institute for Materials Science, 1-1 Namiki, Tsukuba 305-0044, Japan}

\author{Takashi Taniguchi}
\affiliation{Advanced Materials Laboratory, National Institute for Materials Science, 1-1 Namiki, Tsukuba 305-0044, Japan}

\author{Thomas Ihn}
\affiliation{Solid State Physics Laboratory, ETH Z\"{u}rich, 8093 Z\"{u}rich, Switzerland}

\author{Klaus Ensslin}
\affiliation{Solid State Physics Laboratory, ETH Z\"{u}rich, 8093 Z\"{u}rich, Switzerland}

\author{Marcin Mucha-Kruczy\'nski}
\affiliation{Department of Physics, University of Bath, Claverton Down, Bath, BA2 7AY, UK}

\author{Vladimir I. Fal'ko}
\affiliation{Department of Physics, Lancaster University, Lancaster, LA1 4YB, United Kingdom}


\date{\today}

\begin{abstract}
Bilayer graphene is a unique system where both the Fermi energy and the low-energy electron dispersion can be tuned. This is brought about by an interplay between trigonal warping and the band gap opened by a transverse electric field. Here, we drive the Lifshitz transition in bilayer graphene to experimentally controllable carrier densities by applying a large transverse electric field to a h-BN-encapsulated bilayer graphene structure. We perform magneto-transport measurements and investigate the different degeneracies in the Landau level spectrum. At low magnetic fields, the observation of filling factors $-3$ and $-6$ quantum Hall states at low magnetic fields reflects the existence of three maxima at the top of the valence band dispersion. At high magnetic fields all integer quantum Hall states are observed, indicating that deeper in the valence band the constant energy contours are singly-connected. The fact that we observe ferromagnetic quantum Hall states at odd-integer filling factors testifies to the high quality of our sample. This enables us to identify several phase transitions between correlated quantum Hall states at intermediate magnetic fields, in agreement with the calculated evolution of the Landau level spectrum. The observed evolution of the degeneracies therefore reveals the presence of a Lifshitz transition in our system.
\end{abstract}



\maketitle


Fermi surface topology plays an important role in determining the electronic properties of metals \cite{lifshitz}. In bulk metals, the Fermi energy is not easily tunable at the energy scale needed for reaching the Lifshitz transition - a singular point in the band structure where the connectivity of the Fermi surface changes.

\begin{figure}[H]
\includegraphics[width=\columnwidth]{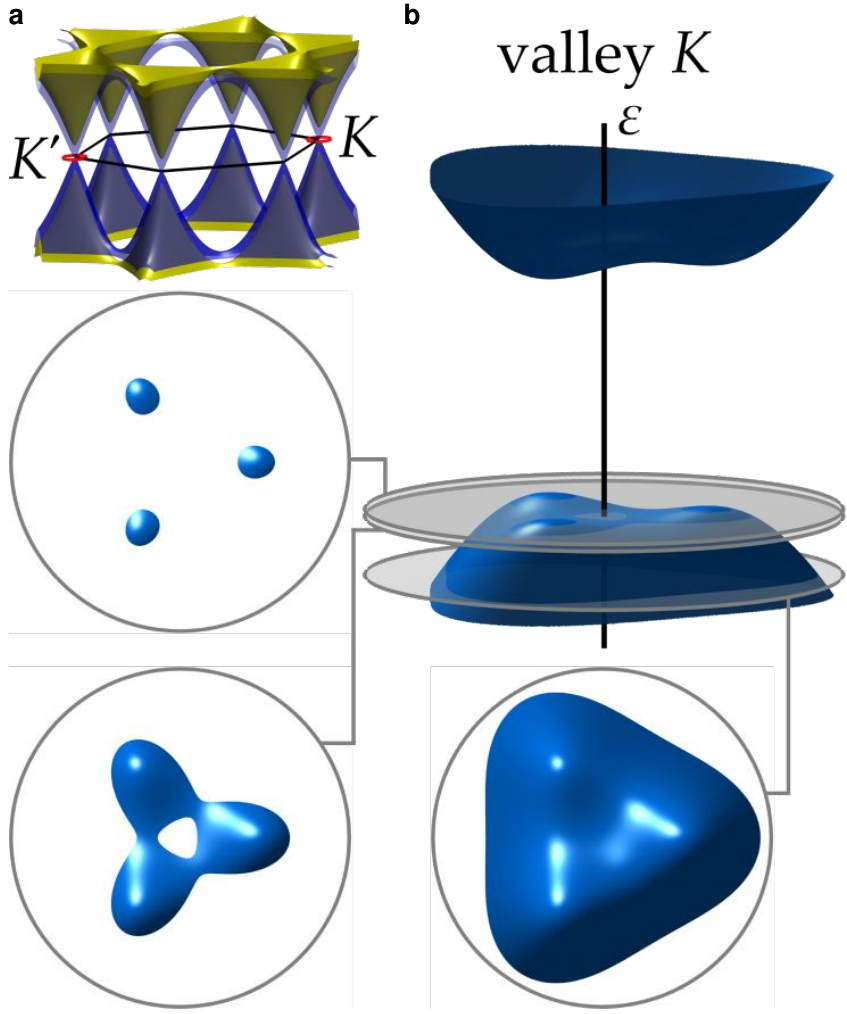}
\caption{(color online). \textbf{Bilayer graphene band structure.} (a) The whole Brillouin zone up to $3~\rm{eV}$ from the neutrality point, including  low-energy bands (blue) and split bands (yellow). (b) Valley K dispersion near the valence band top in gapped BLG. In valleys K and K', the dispersion is inverted as a result of time reversal symmetry. Insets show characteristic cross-sections of  $\epsilon(p)$ discussed in the main text.}
\label{fig1}
\end{figure}

Bilayer graphene (BLG) –- two graphitic layers in Bernal stacking configuration -– is uniquely suitable to control the transformation of the electron spectrum topology at the Fermi energy. Pristine BLG is a gapless semiconductor with almost parabolic conduction and valence bands touching each other in the corners K and K' of its hexagonal Brillouin zone \cite{novoselov_unconventional_2006, mccann_landau-level_2006}, as shown in Fig.~\ref{fig1}(a). Electrons in these bands reside on two sublattices in the neighbouring layers that are not located on top of each other. Very close to the band edges, the BLG dispersion is affected by trigonal warping, enforced by skew interlayer hopping, which produces a Lifshitz transition \cite{lifshitz,mccann_landau-level_2006}. Furthermore, an out-of-plane electric field $E_{z}$ opens an inter-layer asymmetry band gap \cite{mccann_landau-level_2006, mccann_asymmetry_2006, ohta_controlling_2006, castro_biased_2007, oostinga_gate-induced_2008}, with the experimentally measured \cite{ohta_controlling_2006, zhang_direct_2009} gap size $u$, which can be tuned up to $u > 100~\rm{meV}$. The interplay between these two features of BLG results in the electron dispersion illustrated in Fig.~\ref{fig1}(b), where the electron spectrum exhibits a triplet of local band-edge extrema, suggesting an additional three-fold degeneracy of the highest valence band's Landau levels (LLs) at low magnetic fields, on top of the valley and spin degeneracies.

Below, we detect this three-fold degeneracy by performing detailed quantum Hall effect (QHE) studies of BLG sandwiched between hexagonal boron nitride (h-BN). By tuning the gap $u$, we identify additional QHE features related to LLs being affected by the `electron-like' island near the top of the valence band.

\begin{figure}
\includegraphics[width=\columnwidth]{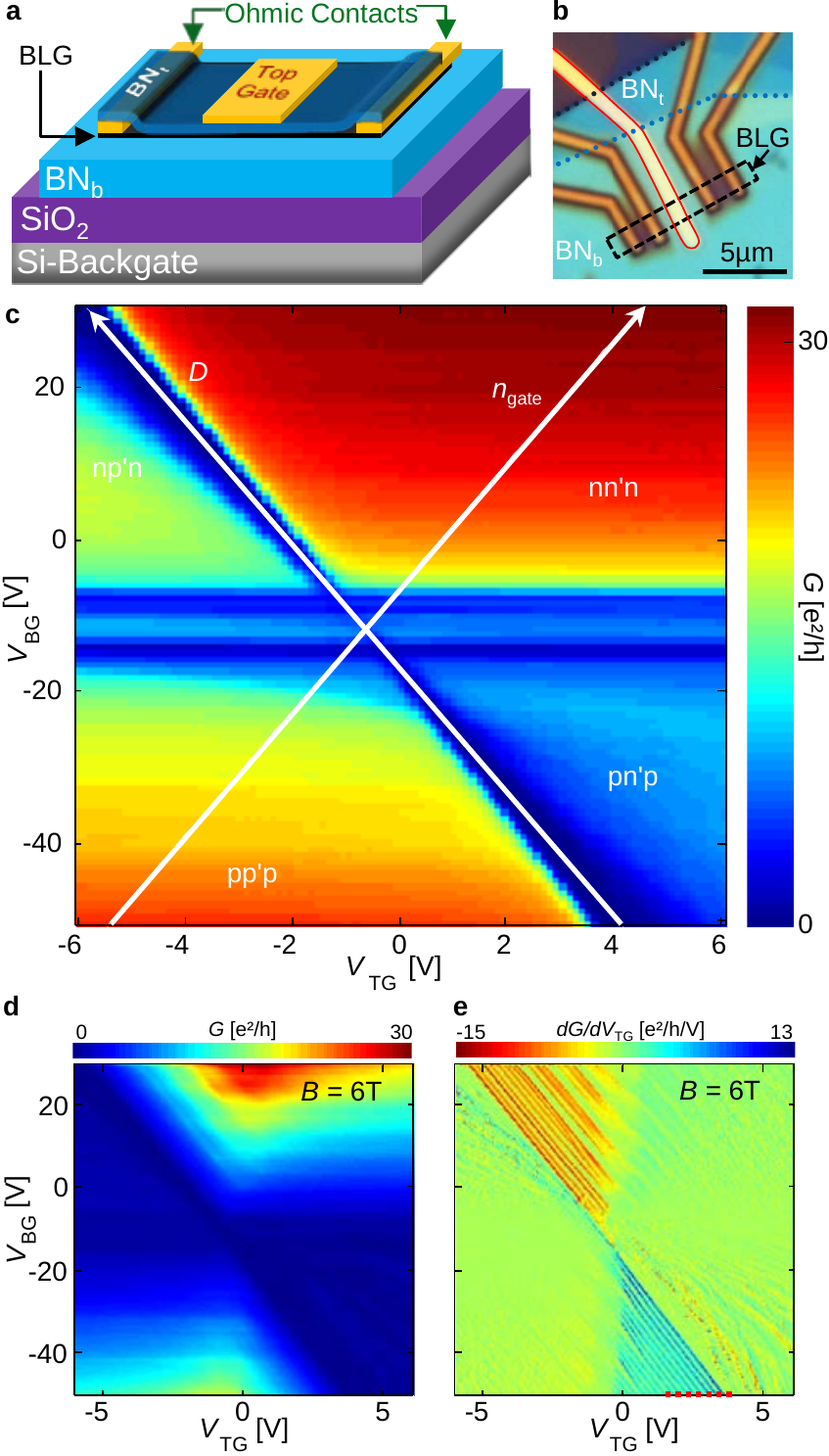}
\caption{(color). \textbf{Characterisation of the device.} (a) Schematics of the device: a bilayer graphene flake is transferred onto the $BN_{b}$ (bottom) flake and contacted. The flake is then covered by another h-BN layer ($BN_{t}$ - top), on top of which a metal top-gate is patterned. (b) Optical microscope image of the device. The ohmic contacts buried under $BN_{t}$ appear dark yellow, while the top-gate is highlighted in red. (c) Conductance map of the device ($B = 0~\rm{T}$, $T = 1.6~\rm{K}$). The displacement field axis and the density axis of the dual-gated area are shown, as well as the four regions of different polarities. At high displacement field, the conductance through the device is suppressed (a gap opens). (d) Conductance map measured at $6~\rm{T}$. (e) Normalised transconductance map at $6~\rm{T}$: in the unipolar cases when $\nu < \nu_{lead}$, all the broken-symmetry states of the area under the top-gate are visible (parallel lines running right or left of the displacement field axis). The red dashed line indicates the cut along which the left Landau fan in Fig.~\ref{fig3}(b) is measured.}
\label{fig2}
\end{figure}

The investigated device is presented in Figs.~\ref{fig2}(a-b): it consists of a $1.3~\rm{\mu m}$-wide BLG stripe contacted with Cr/Au electrodes and sandwiched between two h-BN flakes, stacked using a dry-transfer technique \cite{dean_boron_2010}. To improve surface homogeneity before encapsulation, the BLG was mechanically cleaned with contact-mode AFM \cite{goossens_mechanical_2012}. The resulting device is $3~\rm{\mu m}$-long (distance between inner Ohmic contacts), covered with a $1.1~\rm{\mu m}$-wide top-gate in its centre, and tunable from the back via the degenerately doped Si substrate. The shape of the etched flake is highlighted by the black dotted lines in Fig.~\ref{fig2}(b). The high quality of the device enables us to observe a rich menu of phase transitions investigated as a function of displacement field, magnetic field and carrier density. All measurements were recorded in a two-terminal configuration (inner contacts in Fig.~\ref{fig2}(b)) at a temperature of $1.6~\rm{K}$. The linear conductance $G = I/U$ was obtained by applying a DC bias $U = 100~\rm{\mu V}$ and measuring the current $I$. A small AC top-gate voltage superimposed on the DC voltage $V_\mathrm{TG}$ was used to measure the normalised transconductance $dG/dV_\mathrm{TG}$.

Fig.~\ref{fig2}(c) shows $G$ as a function of back- and top-gate voltages, $V_\mathrm{BG}$ and $V_\mathrm{TG}$. Two horizontal dark blue stripes are visible, indicating features that do not depend on the top-gate voltage: they correspond to charge neutrality in the two only back-gated regions, differing slightly in carrier density. The diagonal blue line relates to the conductance of a sample region which depends both on top- and back-gate voltages: it is the charge neutrality line of the dual-gated region. This line defines the axis of the displacement field, $D$, along which the voltage-induced asymmetry between top and bottom layer of the BLG flake is changed. Along the direction of the $n_\mathrm{gate}$-axis, the density in the dual-gated region can be independently changed at constant $D$. These two axes cross at ($V_\mathrm{BG}^{(0)}$, $V_\mathrm{TG}^{(0)}$) and the dark blue regions of low conductance define four quadrants corresponding to different combinations of carrier polarities in/out-side the dual-gated part. Increasing $D$ from ($V_\mathrm{BG}^{(0)}$, $V_\mathrm{TG}^{(0)}$) decreases the conductance by opening a band gap $u$, with larger $u$ widening the insulating (blue) region. In the following, we relate $u$ and $D$ using the self-consistent calculation \cite{mucha-kruczynski_influence_2009} described in the Supplemental Material \footnote{See Supplemental Material, which includes Ref.\cite{mucha-kruczynski_spectral_2009}}, since hopping via localised states \cite{oostinga_gate-induced_2008,russo_double-gated_2009,weitz_broken-symmetry_2010} obscures a direct measurement of the gap using Arrhenius plots for the conductance.

The high quality of the device is revealed by its behaviour at strong magnetic fields $B$. As found in QHE studies, both in two-terminal monolayer \cite{williams_quantum_2007,ozyilmaz_electronic_2007,amet_gate_2013} and BLG \cite{jing_quantum_2010} devices, the quantized conductance, shown for our device in Fig.~\ref{fig2}(d), is described by the Landauer-B\"{u}ttiker formalism. In the unipolar case ($nn'n$ or $pp'p$) with smaller density in the dual-gated area ($\nu < \nu_\mathrm{lead}$), the outer regions behave as leads and the conductance is given by the number $|\nu|$ of edge modes transmitted through the dual-gated region: $G = (e^{2}/h) |\nu|$. In the $pp'p$ corner of Fig.~\ref{fig2}(d), all integer values of quantized conductance, up to $|\nu| = 16$, are observed, indicating broken valley and spin degeneracy, and the formation of incompressible ferromagnetic QHE states \cite{nomura_quantum_2006, giesbers_gap_2009, feldman_broken-symmetry_2009, young_spin_2012, jr_transport_2012} at odd values of $|\nu|$. These steps are better visible in the normalised transconductance signal shown in Fig.~\ref{fig2}(e) (green regions, with $dG/dV_\mathrm{TG} \rightarrow 0$, separated by blue lines running parallel to the $D$-field axis). In the following we trace the incompressible QHE states using transconductance maps.

\begin{figure}
\includegraphics[width=\columnwidth]{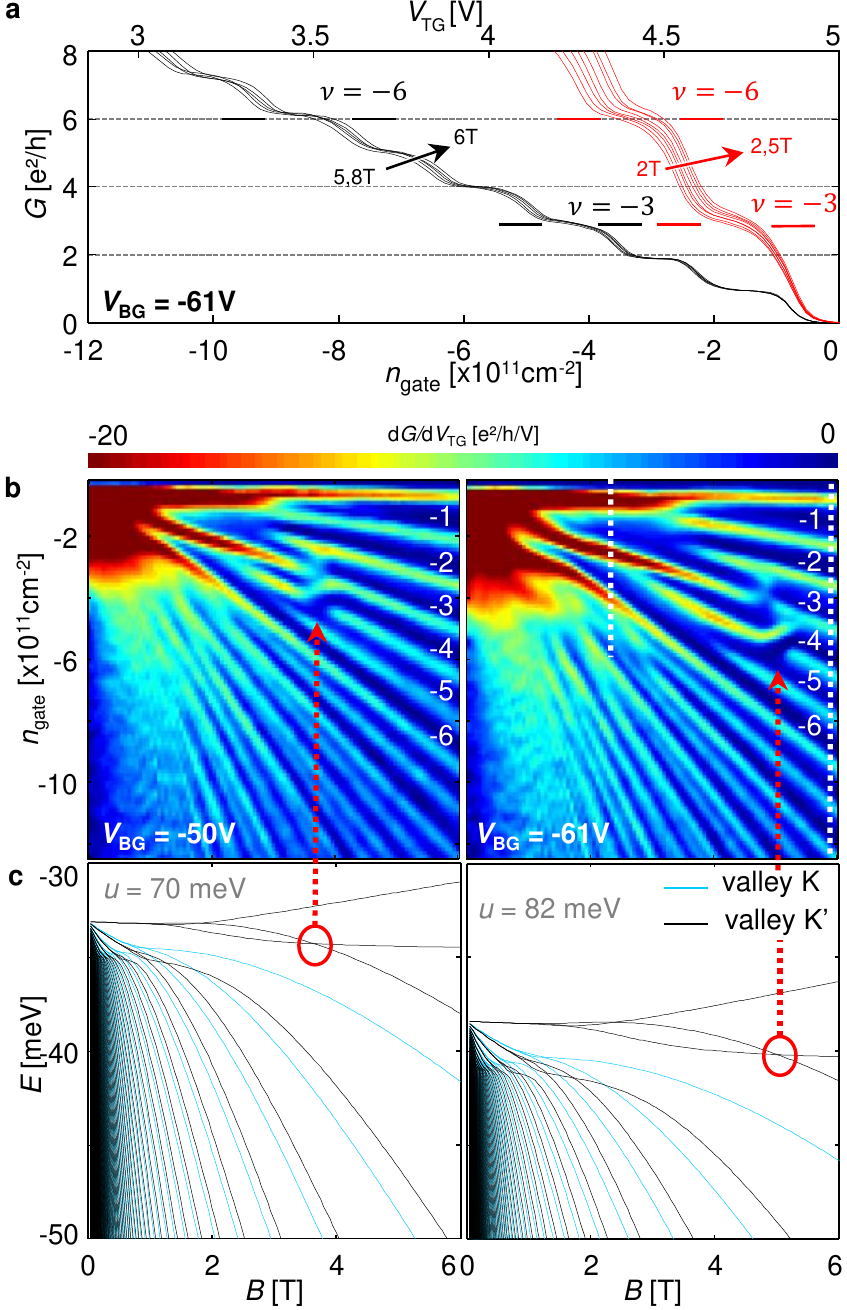}
\caption{(color). \textbf{Magneto-transport through dual-gated bilayer graphene.} (a) Conductance cuts taken along the white dotted lines in (b) (right panel), where $n_\mathrm{gate}$ is varied with $V_\mathrm{TG}$. A contact resistance $R_{C} = 850~\rm{\Omega}$ has been subtracted. At high magnetic field (black curves), all the broken-symmetry states are visible. At lower magnetic fields (red curves), the first conductance plateaus to appear correspond to filling factors $-3$ and $-6$. (b) Normalised transconductance LL spectra measured at $V_\mathrm{BG} = -50~\rm{V}$ ($D = -0.95~\rm{V.nm^{-1}}$ at $n_\mathrm{gate} = 0$), along the red dashed line shown in Fig.~\ref{fig2}(e), and $V_\mathrm{BG} = -61~\rm{V}$ ($D = -1.14~\rm{V.nm^{-1}}$ at at $n_\mathrm{gate} = 0$). Corresponding filling factors are labelled. We observe a crossing between filling factors $-4$ and $-5$. (c) Calculated LL spectra corresponding to gap sizes $u = 70~\rm{meV}$ and $u = 82~\rm{meV}$. Blue (black) lines refer to valley K (K').}
\label{fig3}
\end{figure}

Figure \ref{fig3} shows the evolution of the QHE states in $p$-doped BLG, as a function of the $B$-field. In Fig.~\ref{fig3}(a), we show high $D$-field data ($V_\mathrm{BG} = -61~\rm{V}$) and compare the staircase of quantized conductance steps at high and low $B$-fields ($B \simeq 5.9~\rm{T}$ and $B \simeq 2.25~\rm{T}$). Most strikingly the low-field data shows only robust $\nu = -3$ and $\nu = -6$ states, whereas the high-field data exhibits all integer QHE states. These QHE states are investigated in Fig.~\ref{fig3}(b) as a function of $B$-field for two different ranges of the interlayer asymmetry (tuned with $D$, via $V_\mathrm{BG}$): these plots are recorded sweeping the top-gate voltage at constant $V_\mathrm{BG}$ (see red dashed line in Fig.~\ref{fig2}(e)). Here, blue regions ($dG/dV_\mathrm{TG} \rightarrow 0$) trace incompressible QHE states, with the $\nu = -3$ and $-6$ states persistent down to $B \simeq 1~\rm{T}$, in contrast to other QHE states which disappear at $B \simeq 3~\rm{T}$. In addition, several other features interrupt the $\nu = -3$, $-4$, and $-5$ QHE states.

To interpret the observed evolution of QHE states, we analyse the LL spectrum of our device. We use the $4$-band Hamiltonian \cite{mccann_landau-level_2006}, describing a broad spectral range ($>1~\rm{eV}$) in valleys K ($\zeta = 1$) and K' ($\zeta = -1$),

\begin{equation}
\label{eq1}
H =
\begin{bmatrix}
  \frac{\zeta}{2}u & v_{3} \pi & 0 & v \pi^{\dag} \\
  v_{3} \pi^{\dag} & -\frac{\zeta}{2}u & v \pi & 0 \\
  0 & v \pi^{\dag} & -\frac{\zeta}{2}u & \zeta \gamma_{1} \\
  v \pi & 0 & \zeta \gamma_{1} & \frac{\zeta}{2}u
\end{bmatrix}
~,~
\left\{
\begin{array}{ll}
    \pi = p_{x} + ip_{y} \\
    v = \frac{\sqrt{3}}{2} \frac{a\gamma}{\hbar} \sim 10^{8} cm/s \\
    v_{3} = \frac{\sqrt{3}}{2} \frac{a\gamma_{3}}{\hbar} \sim 0.1v
\end{array}
\right.
\end{equation}

Here, we characterise the electron states using amplitudes $A$ and $B$ on sublattices in layers $1$ and $2$: $4$-spinors ($A_{1}$, $B_{2}$, $A_{2}$, $B_{1}$) and ($B_{2}$, $-A_{1}$, $B_{1}$, $-A_{2}$) in valleys K and K' are basis vectors; $a$ is the graphene lattice constant; vertical and skew interlayer hopping parameters are $\gamma_1  \sim 0.4~\rm{eV}$ and $\gamma_{3}  \sim 0.3~\rm{eV}$, respectively; $\pi = p e^{i\theta}$ and $\pi^{\dag} = pe^{-i\theta}$, where $\theta$ identifies the azimuthal direction of the electron valley momentum $\vec{p}$. This results in the low-energy dispersion,

\begin{equation}
\label{eq2}
\epsilon^{2} \approx \frac{u^{2}}{4} - \left[\left(\frac{u v^{2}}{\gamma_{1}}\right)^2 - {v_{3}}^{2}\right] p^{2} - 2\zeta \frac{v^{2}v_{3}}{\gamma_{1}} p^{3}\cos{3\theta} + \frac{v^{4}}{{\gamma_{1}}^{2}} p^{4},
\end{equation}

featuring three local band-edge extrema near the top of the valence band, Fig.~\ref{fig1}(b), with momentum separation $\delta p \approx \frac{u}{v\sqrt{2}}$ and depth $\delta\epsilon \approx \frac{u^{2}\gamma_{3}}{\sqrt{2}\gamma_{1}\gamma}$. Upon filling BLG with holes, one starts with three separate Fermi `lakes', which for increasing density merge together into a Fermi sea with an island surrounding a local minimum exactly in the centre of the valley. After lowering the energy cut by an additional $\delta\epsilon' \approx \frac{1}{4} \frac{u^{3}}{{\gamma_{1}}^{2}}$, this `electron-like' central patch disappears.

At a finite $B$-field, $\vec{p} =-i\hbar\nabla + e\vec{A}$ (where $\nabla\times\vec{A} = B$) and $\pi$ and $\pi^{\dag}$ become lowering/raising operators in the space of Landau functions $\psi_{n}$. Without an interlayer asymmetry gap, the LL spectrum has two degenerate levels at zero energy~\cite{novoselov_unconventional_2006, mccann_landau-level_2006}, $n = 0$ and $n = 1$, with wave functions located on sublattices $A_{1}$ in valley K and $B_{2}$ in valley K'. In gapped BLG, the sublattice segregation of $n = 0$ and $n = 1$ levels in the K and K' valleys lifts the valley degeneracy, as shown in Fig.~\ref{fig3}(c), where we compare the calculated spectra in the two valleys (for details, see section $2$ in Supplemental Material \cite{Note1}). Assuming spin degeneracy, the characteristic spectra are calculated and exemplified in Fig.~\ref{fig3}(c), using $u = 70~\rm{meV}$ and $82~\rm{meV}$. The choice of $u$ in Fig.~\ref{fig3} results from the estimates of the displacement field $D$ obtained from parallel plate capacitor model and a self-consistent calculation of $u$ as a function of $D$ (see \cite{Note1} for details). The exact numerical values were then chosen to match best the experimental data. The resulting spectra show a three-fold degenerate LL at low B-fields (hence six-fold including spin degeneracy), originating from the triplet of valence band extrema. This degeneracy persists over the same $B$-field interval as the $\nu = -6$ state in the measurements shown in Fig.~\ref{fig3}(b). Exchange interaction between electrons in a three-fold degenerate LL leads to the formation of a $\nu  = -3$ ferromagnetic QHE state, observed within the same low-field interval.

At higher $B$-field, the triplet of the lowest LLs splits by mixing of the momentum space regions near the valence band top, especially near the Lifshitz transition in the BLG energy spectrum. Hence, at higher field, one would expect the appearance of all integer (spin non-polarised/even and ferromagnetic/odd) QHE states, except for the $B$-fields near $B_{\ast}$ where the $2^\mathrm{nd}$ and $3^\mathrm{rd}$ LLs cross [Fig.~\ref{fig3}(c)].  This unusual crossing results from the existence of the central electron-like island in the constant energy contour at intermediate energies, disconnected from the outer hole-like dispersion branch [bottom left inset in Fig.~\ref{fig1}(b)]: these two parts of momentum space support LLs that do not mix/repel each other. This occasional double degeneracy of orbital LLs at $B = B_{\ast}$ suppresses the gap of $\nu = -3$ and $-5$ ferromagnetic states in this degenerate pair of LLs. It also suggests a transition of the $\nu = -4$ state, from a non-spin-polarised configuration at $B > B_{\ast}$ and $B < B_{\ast}$ to a spin-polarised state occupying the degenerate pair of LLs at $B \approx B_{\ast}$. As indicated by red arrows in Figs.~\ref{fig3}(b-c), the calculated LL crossings in Fig.~\ref{fig3}(c) correlate with the measured values of the $B$-field where additional features interrupt the incompressible QHE states $\nu = -3$, $-4$ and $-5$ in the transconductance plots in Fig.~\ref{fig3}(b).

The high electronic quality of the mechanically-cleaned BLG and the electrical robustness of the h-BN enable us to change the band structure topology.  Most interestingly, we find that the Lifshitz transition, which is a single-particle phenomenon, leads to interaction-driven transitions between quantum Hall ferromagnets, associated with the crossover from the three-fold-degenerate to the non-degenerate orbital LL spectrum. For the future, a larger electrically controllable band gap in BLG would enable us to investigate in detail the influence of the van Hove singularity in the density of states, characteristic for the Lifshitz transition, on quantum transport characteristics, while cleaner samples may offer access to new regimes of correlated QHE states in the regime of three-fold degenerate LLs.

\begin{acknowledgments}
We thank A. Kozikov for discussions. We acknowledge financial support from the Marie Curie ITNs $S^{3}NANO$, QNET, the Swiss National Science Foundation via NCCR Quantum Science and Technology, the ERC Synergy Grant 'Hetero$2$D' and the European Graphene Flagship Project.
\end{acknowledgments}

\bibliography{QHE_Papers}

\end{document}